\documentclass{iet-ell}
\usepackage{algorithmic}
\usepackage{algorithm}
\usepackage{marvosym}
\usepackage{caption}
\usepackage{booktabs}
\usepackage{graphicx}
\usepackage{soul,color}

\begin{document}

% Title
\title{YOLOv9 for Fracture Detection in Pediatric Wrist Trauma X-ray Images}

%Authors, affiliations address.
\author[af1]{Chun-Tse~Chien}
\orcid{0009-0008-7549-4021}
\author[af2]{Rui-Yang~Ju}
\orcid{0000-0003-2240-1377}
\author[af3]{Kuang-Yi~Chou}
\author[af1]{Jen-Shiun~Chiang\textsuperscript{\Letter, }}
\orcid{0000-0001-7536-8967}

\affil[af1]{Department of Electrical and Computer Engineering, Tamkang University, New Taipei City, 251301, Taiwan}
\affil[af2]{Graduate Institute of Networking and Multimedia, National Taiwan University, Taipei City, 106335, Taiwan}
\affil[af3]{School of Nursing, National Taipei University of Nursing and Health Sciences, Taipei City, 112303, Taiwan}

\corresp{\textsuperscript{\Letter}Email: jsken.chiang@gmail.com}

%Abstract
\begin{abstract}%
The introduction of YOLOv9, the latest version of the You Only Look Once (YOLO) series, has led to its widespread adoption across various scenarios. This paper is the first to apply the YOLOv9 algorithm model to the fracture detection task as computer-assisted diagnosis (CAD) to help radiologists and surgeons to interpret X-ray images. Specifically, this paper trained the model on the GRAZPEDWRI-DX dataset and extended the training set using data augmentation techniques to improve the model performance. Experimental results demonstrate that compared to the mAP 50-95 of the current state-of-the-art (SOTA) model, the YOLOv9 model increased the value from 42.16\% to 43.73\%, with an improvement of 3.7\%. The implementation code is publicly available at \url{https://github.com/RuiyangJu/YOLOv9-Fracture-Detection}.
\end{abstract}

\maketitle
\section{Introduction}
The accurate interpretation of X-ray images is critical to the success of surgery. However, there is a shortage of experienced radiologists in some areas, which poses a challenge \cite{rimmer2017radiologist,burki2018shortfall}. Computer-assisted diagnosis (CAD) helps specialists such as radiologists and surgeons to interpret medical images, including magnetic resonance imaging (MRI), computed tomography (CT), and X-ray images. However, it cannot be relied on as the sole means of detection due to uncertainties in its prediction accuracy. With the rapid development of deep learning in computer vision tasks \cite{sezer2021detection,yaug2022artificial}, its application to medical image processing has obtained satisfactory results \cite{adams2021artificial}.

You Only Look Once (YOLO) series \cite{redmon2016you,bochkovskiy2020yolov4,glenn2023,wang2024yolov9} are the main neural networks for real-time object detection task, widely employed in fracture detection \cite{hrvzic2022fracture,ju2023fracture,chien2024yolov8}. Wrist fractures in children are more common cases and the GRAZPEDWRI-DX dataset \cite{nagy2022pediatric} provides 20,327 X-ray images of pediatric wrist trauma that can be used in fracture detection tasks. Ju \emph{et al.} \cite{ju2023fracture} first used the YOLOv8 \cite{glenn2023} model for fracture detection on this dataset. Since attention mechanisms \cite{hu2018squeeze,woo2018cbam,wang2020eca,zhang2021sa,liu2021global} have excellent results in enhancing the performance of neural network models, Chien \emph{et al.} \cite{chien2024yolov8} achieved the state-of-the-art (SOTA) performance by incorporating different attention mechanisms into the YOLOv8 model.

\begin{figure}[ht]
\centering
\includegraphics[scale=0.5]{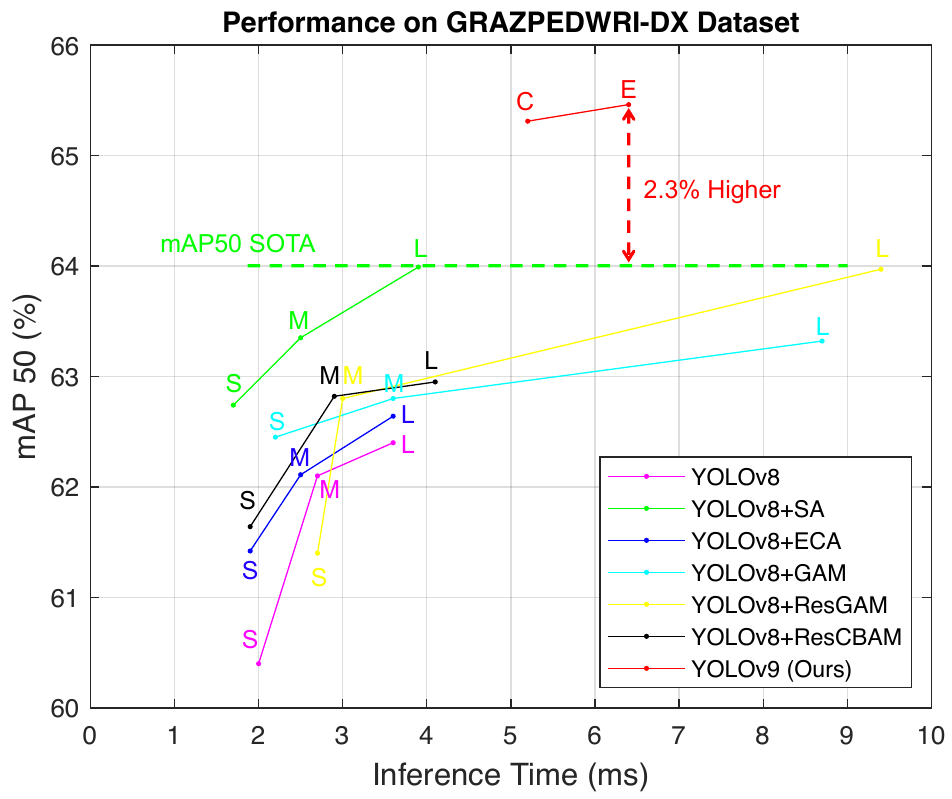}
\caption{Comparisons of fracture detection models on the GRAZPEDWRI-DX dataset when the input image size is 640. In terms of accuracy, our models outperform all previous models at the state-of-the-art level.}
\label{fig:1}
\vspace{-3mm}
\end{figure}

With the presentation of YOLOv9 \cite{wang2024yolov9}, which achieved remarkable model performance on the MS COCO 2017 \cite{lin2014microsoft} benchmark dataset, achieving an impressive increase in mAP from 53.9\% in YOLOv8 to 55.6\%. This paper first trained the YOLOv9 model on the GRAZPEDWRI-DX dataset and obtained the SOTA performance, as shown in Fig \ref{fig:1}.

The main contributions of this paper are as follows:
\vspace{-3mm}
\begin{enumerate}
    \item This paper is the first to apply YOLOv9 to the fracture detection task, demonstrating that the model not only has the excellent performance in real-time object detection across real-life scenarios, but also has good results in medical image recognition.
    \item This paper addresses the issue of information loss in fracture detection on X-ray images by employing YOLOv9 algorithm, and aims to preserve more information during model training on low-features X-ray images, enhancing the performance of the model.
    \item The mAP 50-95 of YOLOv9 model trained on the GRAZPEDWRI-DX dataset is significantly improved, achieving the SOTA level.
\vspace{-3mm}
\end{enumerate}

\section{Related Works}
In the field of object detection task, detectors typically employ either one-stage or two-stage algorithms. Compared to two-stage object detectors, YOLO series models offer a balanced combination of accuracy and inference speed, making them suitable for deployment on mobile computing platforms for medical image recognition. Hržić \emph{et al.} \cite{hrvzic2022fracture} employed the YOLOv4 \cite{bochkovskiy2020yolov4} model for fracture detection on the GRAZPEDWRI-DX dataset \cite{nagy2022pediatric}, which was the first to demonstrate that the models of YOLO series can assist radiologists in more accurately predicting wrist injuries in children on X-ray images. Ju \emph{et al.} \cite{ju2023fracture} developed the ``Fracture Detection Using YOLOv8 App'' to aid surgeons in interpreting X-ray images for fractures, aiming to reduce misclassification and enhance the information available for fracture surgery. Chien \emph{et al.} \cite{chien2024yolov8} proposed the YOLOv8-AM model, incorporating four attention modules \cite{woo2018cbam,wang2020eca,zhang2021sa,liu2021global}, and trained them on the GRAZPEDWRI-DX dataset \cite{nagy2022pediatric}. Specifically, the ResCBAM-based YOLOv8-AM model achieved a significant improvement in mAP 50 value, increasing from 63.6\% to 65.8\% compared to the YOLOv8 model, which reached the performance of SOTA. While the application of the models of YOLO series to medical image recognition is a hot research topic, to date, no one has utilized YOLOv9 \cite{wang2024yolov9} for fracture detection.

\begin{figure*}[ht]
\centering
\includegraphics[scale=0.5]{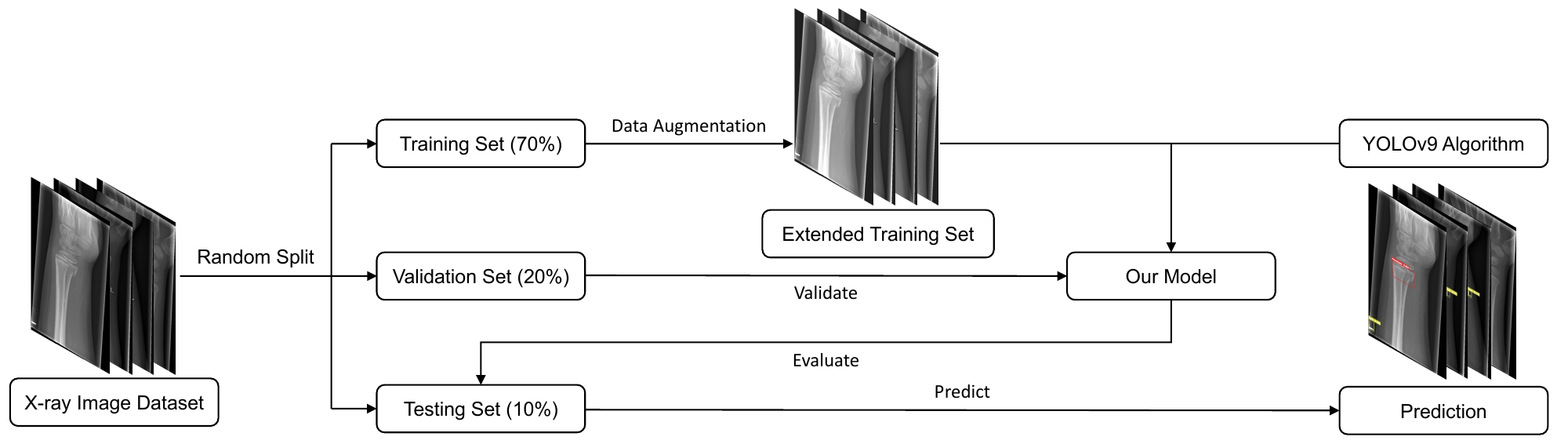}
\caption{An overall flowchart of the YOLOv9 algorithm model applied to the fracture detection task.}
\label{fig:2}
\end{figure*}

\begin{figure*}[ht]
\centering
\includegraphics[scale=0.5]{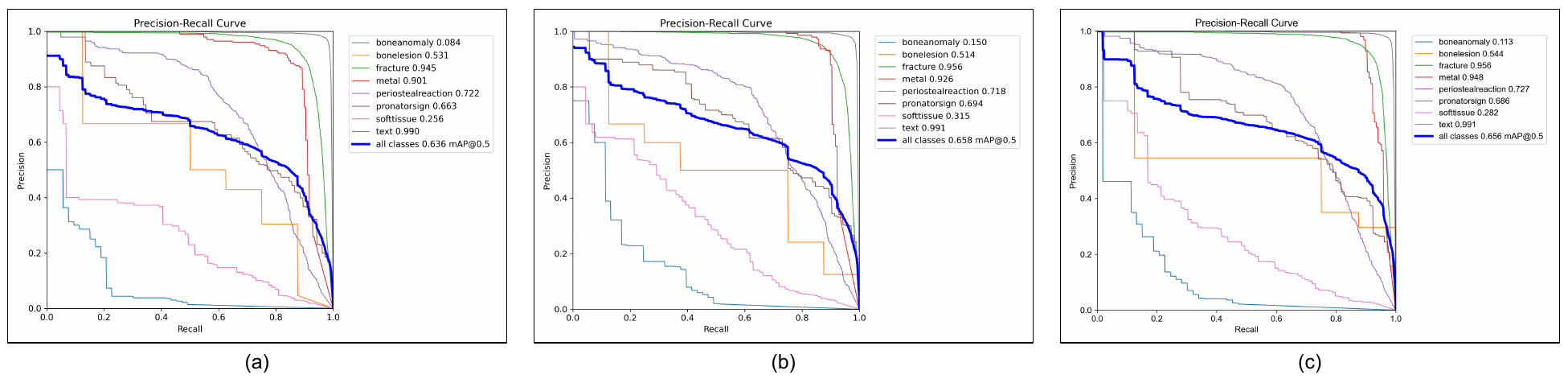}
\caption{Visualization of the accuracy of predicting each class using different models on the GRAZPEDWRI-DX dataset with the input image size of 1024. (a) YOLOv8, (b) YOLOv8+ResCBAM (SOTA), (c) YOLOv9 (Ours).}
\label{fig:3}
\vspace{-4mm}
\end{figure*}

\section{Method}
\subsection{YOLOv9}
\vspace{-3mm}
Neural networks often have the challenge of information loss since the input data undergoes multiple layers of feature extraction and spatial transformation, resulting in the loss of the original information. This issue is particularly pronounced in X-ray images, where the low-features present significant difficult in fracture detection tasks. Specifically, models trained on such low-featured images tend to perform poorly, and addressing the problem of information loss could substantially enhance the accuracy of model predictions. To address this, we utilizing the YOLOv9 algorithm, which leverages the Programmable Gradient Information (PGI) and the Generalized Efficient Layer Aggregation Network (GELAN) to more effectively extract key features. 

\subsubsection{Programmable Gradient Information}
PGI is an auxiliary supervision framework designed to manage the propagation of gradient information across various semantic levels, to improve the detection capability of the model. PGI comprises three main components: main branch, auxiliary reversible branch, and multi-level auxiliary information. During the inference process, it exclusively employs the main branch, which handles both forward and back propagation. As the network becomes deeper, an information bottleneck may occur, leading to loss functions that fail to produce useful gradients. In such cases, auxiliary reversible branch employs reversible functions to preserve information integrity and mitigate information loss in the main branch. Additionally, multi-level auxiliary information addresses the issue of error accumulation from the deep supervision mechanism, improving the learning capacity of the model through the introduction of supplementary information at different levels. Notably, research \cite{wang2024yolov9} highlighted the efficacy of PGI in preserving information during training, particularly in scenarios with limited features. This provides the theoretical basis for the YOLOv9 model to have excellent performance in fracture detection tasks.

\begin{table}[ht]
\caption{Quantitative comparison with other state-of-the-art models for fracture detection on the GRAZPEDWRI-DX datasets when the input image size is 640.}
\label{tab:1}
\setlength{\tabcolsep}{0.8mm}{
\begin{tabular}{lcccccc}
\toprule[1pt]
\textbf{Model} & \textbf{\begin{tabular}[c]{@{}c@{}}Params\\ (M)\end{tabular}} & \textbf{\begin{tabular}[c]{@{}c@{}}FLOPs\\ (G)\end{tabular}} & \textbf{\begin{tabular}[c]{@{}c@{}}F1\\ Score\end{tabular}} & \textbf{\begin{tabular}[c]{@{}c@{}}mAP 50\\ (\%)\end{tabular}} & \textbf{\begin{tabular}[c]{@{}c@{}}mAP 50-95\\ (\%)\end{tabular}} & \textbf{\begin{tabular}[c]{@{}c@{}}Speed\tabnoteref{t1}\\ (ms)\end{tabular}} \\
\midrule[1pt]
YOLOv8\cite{glenn2023} & 43.61 & 164.9 & 0.59 & 62.44 & 40.32 & 3.6 \\
YOLOv8+SA\cite{zhang2021sa} & 43.64 & 165.4 & 0.62 & 63.99 & 41.49 & 3.9 \\
YOLOv8+ECA\cite{wang2020eca} & 43.64 & 165.5 & 0.61 & 62.64 & 40.21 & 3.6 \\ 
YOLOv8+GAM\cite{liu2021global} & 49.29 & 183.5 & 0.60 & 63.32 & 40.74 & 8.7 \\ 
YOLOv8+ResGAM\cite{chien2024yolov8} & 49.29 & 183.5 & 0.62 & 63.97 & 41.18 & 9.4 \\ 
YOLOv8+ResCBAM\cite{woo2018cbam} & 53.87 & 196.2 & 0.62 & 62.95 & 40.10 & 4.1 \\
\midrule
\textbf{YOLOv9-C (Ours)} & 51.02 & 239.0 & 0.64 & 65.31 & 42.66 & 5.2 \\
\textbf{YOLOv9-E (Ours)} & 69.42 & 244.9 & 0.64 & 65.46 & 43.32 & 6.4 \\
\bottomrule[1pt]
\end{tabular}}
\tablenote{\textit{Note:} The model size of all YOLOv8 and its variants listed in the table is large.}
\tablenote[t1]{Speed is the total time for preprocessing, inference, and post-processing.}
\vspace{-4mm}
\end{table}

\subsubsection{Generalized Efficient Layer Aggregation Network}
To enhance information integration and propagation efficiency in model training, YOLOv9 introduced a novel lightweight network architecture named Generalized Efficient Layer Aggregation Network (GELAN). GELAN integrates CSPNet \cite{wang2020cspnet} and ELAN \cite{wang2022designing} to efficiently aggregate network information, reducing information loss in propagation and enhancing inter-layer information interaction. This architecture is particularly suitable for fracture detection in environments with limited computing resources due to its lower parameters and computational complexity.

\begin{table}[ht]
\caption{Quantitative comparison with other state-of-the-art models for fracture detection on the GRAZPEDWRI-DX datasets when the input image size is 1024.}
\label{tab:2}
\setlength{\tabcolsep}{0.8mm}{
\begin{tabular}{lcccccc}
\toprule[1pt]
\textbf{Model} & \textbf{\begin{tabular}[c]{@{}c@{}}Params\\ (M)\end{tabular}} & \textbf{\begin{tabular}[c]{@{}c@{}}FLOPs\\ (G)\end{tabular}} & \textbf{\begin{tabular}[c]{@{}c@{}}F1\\ Score\end{tabular}} & \textbf{\begin{tabular}[c]{@{}c@{}}mAP 50\\ (\%)\end{tabular}} & \textbf{\begin{tabular}[c]{@{}c@{}}mAP 50-95\\ (\%)\end{tabular}} & \textbf{\begin{tabular}[c]{@{}c@{}}Speed\tabnoteref{t2}\\ (ms)\end{tabular}} \\
\midrule[1pt]
YOLOv8\cite{glenn2023} & 43.61 & 164.9 & 0.62 & 63.63 & 40.41 & 7.7 \\
YOLOv8+SA\cite{zhang2021sa} & 43.64 & 165.4 & 0.63 & 64.25 & 41.64 & 8.0 \\
YOLOv8+ECA\cite{wang2020eca} & 43.64 & 165.5 & 0.65 & 64.26 & 41.94 & 7.7 \\ 
YOLOv8+GAM\cite{liu2021global} & 49.29 & 183.5 & 0.65 & 64.26 & 41.00 & 12.7 \\ 
YOLOv8+ResGAM\cite{chien2024yolov8} & 49.29 & 183.5 & 0.64 & 64.98 & 41.75 & 18.1 \\
YOLOv8+ResCBAM\cite{woo2018cbam} & 53.87 & 196.2 & 0.64 & 65.78 & 42.16 & 8.7 \\
\midrule
\textbf{YOLOv9-C (Ours)} & 51.02 & 239.0 & 0.66 & 65.57 & 43.70 & 12.7 \\
\textbf{YOLOv9-E (Ours)} & 69.42 & 244.9 & 0.66 & 65.62 & 43.73 & 16.1 \\
\bottomrule[1pt]
\end{tabular}}
\tablenote{\textit{Note:} The model size of all YOLOv8 and its variants listed in the table is large.}
\tablenote[t2]{Speed is the total time for preprocessing, inference, and post-processing.}
\vspace{-4mm}
\end{table}

\subsection{Data Processing and Augmentation}
Fig \ref{fig:2} illustrates the flowchart of the experiments conducted in this study. Since the publisher of the GRAZPEDWRI-DX \cite{nagy2022pediatric} dataset did not provide predefined training, validation, and test sets, we randomly assigned 70\% to training set, 20\% to validation set, and 10\% to test set during the data processing. Moreover, due to the limited brightness diversity of low-featured X-ray images, models trained only on these images may not generalize well to X-ray images in other environments. To enhance the robustness \cite{ansari2022lightweight,mohanty2022toward,ansari2023dense} of the model, we followed the data augmentation technique used by Ju \emph{et al.} \cite{ju2023fracture} to extend the training set. Specifically, we fine-tuned the contrast and luminance of the X-ray images using the addWeighted function from the OpenCV library.

\section{Experiment}
\subsection{Dataset}
\vspace{-3mm}
GRAZPEDWRI-DX \cite{nagy2022pediatric} is a public dataset provided by the Medical University of Graz, which contains 20,327 X-ray images of pediatric wrist trauma. These X-ray images were collected by a team of pediatric radiologists at the University Hospital Graz from 2008 to 2018. The dataset comprises 6,091 patients and 10,643 studies, with a total of 74,459 labeled images, representing 67,771 labeled objects. There is a significant class imbalance in this dataset. Specifically, the top two labeled objects are ``text'' (23,722 instances) and ``fracture'' (18,090 instances), whereas the number of instances for ``bone anomaly'' and ``soft tissue'' are only 276 and 464, respectively. All models use the same training data.

\subsection{Experiment setup}
The experiments in this paper utilized one single NVIDIA GeForce RTX 3090 GPU, employing Python with the PyTorch framework. Before training our model, we employed the YOLOv9 model weights pretrained on the MS COCO 2017 \cite{lin2014microsoft} dataset. In the training process, we trained the model using the SGD optimizer, with a weight decay rate set to 5e-4 and a momentum of 0.937. We followed the research \cite{ju2023fracture} to set the initial learning rate to 1e-2, the number of epochs to 100. Due to resource limitations (24GB memory) imposed by a single GPU, a batch size of 16 was employed for training the model.

\subsection{Experimental Results}
It is well known that the YOLO series models are designed for real-time object detection and are suitable to be deployed in web or mobile applications to assist surgeons in pediatric wrist fracture diagnosis in all regions of the world. The research \cite{nagy2022pediatric} has shown that the GRAZPEDWRI-DX dataset is the first publicly available and pertinent pediatric dataset, so we evaluate the performance of different models on it.

To evaluate the performance of YOLOv9 and other SOTA models in real diagnostic scenarios, this study compares model size (parameters and floating point operations), accuracy (F1 score, mean average precision at 50\% (mAP 50), and mean average precision from 50\% to 95\% (mAP 50-95)), and inference time. It is widely recognized that using larger input image sizes improves prediction accuracy but also requires more computational resources. Therefore, we conducted two experiments with input image sizes of 640 and 1024 for various scenarios, and the results are presented in Tables \ref{tab:1} and \ref{tab:2}. With the input size of 640, both YOLOv9-C (Compact) and YOLOv9-E (Extended) demonstrate significantly improved mAP, while maintaining a reasonable inference speed. Specifically, YOLOv9-E achieves mAP 50-95 of 43.32\%, which is 4.4\% higher than 41.49\% achieved by the current SOTA model YOLOv8+SA. When the input image size is 1024, the mAP 50-95 of YOLOv9-E reaches 43.73\%, which also obtains the SOTA performance. However, due to the increased inference time, it is more suitable for deployment on devices with high computing resources. In summary, the YOLOv9 model effectively balances inference time with fracture detection accuracy, making it suitable for deployment in web applications as a CAD system to assist surgeons in analyzing X-ray images.

To facilitate further performance enhancement, we visualized the accuracy of various models in predicting each class, as illustrated in Fig. \ref{fig:3}. Across all models, the ability to accurately predict ``fracture'', ``metal'', and ``text'' classes exceeded 90\%, with YOLOv9 notably achieving a 94.8\% accuracy in ``metal'' class prediction. However, the performance in predicting bone anomalies and soft tissue classes was poorer, with YOLOv9 achieving only 11.3\% and 28.2\% accuracy, respectively. This significantly impacted the average accuracy of YOLOv9 in fracture detection.

\section{Discussion}
The performance of YOLO series models is significantly influenced by the quality and diversity of the training datasets. As mentioned above, the limited ability of these methods to accurately predict ``bone anomaly'' and ``soft tissue'' classes is due to the insufficient number of relevant data in the GRAZPEDWRI-DX dataset. Therefore, a future research can enhance performance by improving the models with additional information on ``bone anomaly'' and ``soft tissue'' classes.

This study focuses exclusively on the detection of pediatric wrist fractures, limiting its applicability to fractures in other body regions. To enhance the model's utility across new cases, we have made the trained weight files available on our GitHub. Users can leverage these files to train the model using X-ray images of different fracture types, enabling the model to achieve reliable predictions with minimal training data.

\section{Conclusion}
The models of YOLO series can serve as CAD to assist radiologists and surgeons in interpreting X-ray images. However, the predictions from the previous models are often unsatisfactory due to the low features of X-ray images. This paper first introduces the application of YOLOv9 to fracture detection, addressing the issue of information loss during model training by employing the newly proposed PGI and GELAN. Experimental results indicate that the YOLOv9 model achieves SOTA performance on the GRAZPEDWRI-DX dataset, proving the effectiveness of this method.

\begin{acks}
This research is supported by National Science and Technology Council of Taiwan, under Grant Number: NSTC 112-2221-E-032-037-MY2.
\end{acks}

\bibliography{iet-ell}
\bibliographystyle{iet}
\end{document}